\documentclass{mn2e}
\usepackage{epsf,times}
\usepackage{amsmath}
\newcommand{\etal}{{et al}\/.}

\begin{document}
\title[TeV jets in Cen A \& M87]{Modelling TeV $\gamma$-ray emission from the
  kiloparsec-scale jets of Centaurus~A and M87}
\author[M.J.~Hardcastle \& J.H.~Croston]{M.J.\ Hardcastle$^1$\thanks{E-mail:
    m.j.hardcastle@herts.ac.uk} and J.H.\ Croston$^2$\\$^1$ School of Physics,
  Astronomy and Mathematics, University of
Hertfordshire, College Lane, Hatfield, Hertfordshire AL10 9AB\\
$^2$ School of Physics and Astronomy, University of Southampton,
Southampton, SO17 1BJ\\}
\maketitle
\begin{abstract}
The widespread detection of synchrotron X-ray emission from the jets
of low-power, nearby radio galaxies implies the presence of electrons
at and above TeV energies. In this paper we explore the possibility
that the TeV $\gamma$-rays detected from the radio galaxies Cen A and
M87, which both have bright, well-studied X-ray jets, are produced at
least in part by inverse-Compton scattering of various photon fields
by the high-energy electrons responsible for the synchrotron X-rays on
kiloparsec scales. We describe a new numerical code that we have
developed to carry out inverse-Compton calculations taking account of
all the relevant physics and using detailed models of the jets and the
photon fields in which they are embedded, and show that existing
constraints on the very high-energy (VHE) $\gamma$-ray fluxes of these
two objects already place significant constraints on the magnetic
field strengths in the jet in Cen A. Finally, we discuss the prospects
for constraints on radio galaxy jet physics that may be obtained from
observations with the Cerenkov Telescope Array (CTA).
\end{abstract}
\begin{keywords}
galaxies: active -- gamma-rays: observations -- radiation mechanisms: non-thermal
\end{keywords}

\section{Introduction}

The existence of X-ray counterparts to the bright radio jets in the
nearby radio galaxies Centaurus A and M87 has been known for many
years (Feigelson \etal\ 1981; Biretta, Stern \& Harris 1991). More
recently, it has become clear that such X-ray jets are common,
possibly ubiquitous (Hardcastle \etal 2001; Worrall \etal\ 2001) in
classical low-power, Fanaroff \& Riley (1974) class I (FRI) radio
galaxies. From the connection to the radio through infrared, optical
and ultraviolet spectrum, it appears that the X-ray emission is
synchrotron in origin (Hardcastle \etal 2001; Hardcastle, Kraft \&
Worrall 2006); for plausible (equipartition) values of the magnetic
field strength in the jet, this means that we are seeing emission from
relativistic electrons with TeV energies. Because of the synchrotron
loss timescales of these electrons in the equipartition magnetic
fields, which are of the order of tens to hundreds of years, it is
widely accepted that the X-ray detections of well-resolved, kpc-scale
emission from FRI jets requires an in situ acceleration process, and
observations of Cen A and M87, the best-resolved and best-studied of
the X-ray jets, have been used to argue both for localized particle
acceleration at shocks and for a more distributed acceleration process
(Hardcastle \etal\ 2003; Perlman \& Wilson 2005; Goodger \etal\ 2010).

However, these arguments are based on assumptions about the magnetic
field strengths in the jets. Another consequence of the availability
of sensitive X-ray observations of radio galaxies has been the
widespread detection of X-ray inverse-Compton emission from the lobes
and hotspots of the more powerful FRII radio galaxies (e.g. Harris
\etal\ 1994; Hardcastle \etal\ 2002; Isobe \etal\ 2002; Hardcastle
\etal\ 2004; Kataoka \& Stawarz 2005; Croston \etal\ 2005) which has
allowed us to show that the field strengths in those regions of radio
galaxies are generally within a factor of a few of the conventional
equipartition value, assuming no energetically dominant proton
population. But in FRI jets, the strong synchrotron X-ray emission
means that we cannot use inverse-Compton X-rays to measure
magnetic field strength; indeed, with the exception
of the recent Fermi measurement of magnetic field strengths in the
giant lobes of Cen A (Abdo \etal\ 2010) we have no direct
inverse-Compton constraints on the magnetic field strength of {\it
  any} components of the FRI population. Among other things, this
restricts our ability to draw conclusions about the particle
acceleration properties of FRI jets.

The inference from the X-ray observations that very-high-energy
leptons are present in the FRI jets does, however, imply that
inverse-Compton emission from these regions should be detectable up to
very high energies: electrons with energies in the 1-10
TeV energy range can inverse-Compton scatter a suitable photon
population into the TeV $\gamma$-ray band. FRI jets are in principle
well supplied with parent photon populations. In addition to the
cosmic microwave background, which is the principal photon field
involved in the inverse-Compton detection of radio galaxy lobes, and
the extragalactic background light (EBL), which is also always
present, synchrotron self-Compton emission (SSC) may be important due
to the relative compactness and brightness of FRI jets, and they are
also exposed to any emission that may be visible to them (though
usually not directly to us) from the active nucleus itself and from
the parsec-scale jet, which will normally be beamed in the direction
of the kpc-scale jet. Finally, in the centres of elliptical galaxies,
the energetically dominant photons are starlight: the energy density
in starlight at the centre of a large elliptical galaxy can be $\sim 2
\times 10^{-12}$ J m$^{-3}$, exceeding that in the $z=0$ CMB by nearly
two orders of magnitude. This fact led Stawarz, Sikora \& Ostrowski (2003)
to predict that the inverse-Compton scattering of starlight would
provide an important contribution to the emission of FRI jets in the
TeV regime.

The opportunity to confront these inverse-Compton predictions has been
provided by the detection of M87 and, recently, Cen A with Cerenkov
imaging telescopes sensitive in the TeV range (Aharonian \etal\ 2003;
Aharonian \etal\ 2009). The spatial resolution of the current
generation of these instruments is not sufficient to distinguish
between emission from the kpc-scale jet and emission from the active
nucleus or from the jet on smaller scales; from the widespread
detection of, often strongly variable, TeV emission from blazars we
know that small-scale jets are certainly capable of producing
high-energy $\gamma$-rays, while the detection of short-timescale
variability in the TeV emission from M87 (e.g.\ Acciari \etal\ 2009,
2010) is evidence that some at least of this emission originates on
small spatial scales, either in the pc-scale jet or possibly in the
peculiar jet knot HST-1 (Cheung, Harris \& Stawarz 2007). However,
even if we treat the level of TeV emission (or, in the case of M87,
its non-variable component) as giving us an upper limit on the
inverse-Compton emission from the kpc-scale jets, these detections
give us a lower limit on the magnetic field strength that cannot be
obtained in any other way, so long as we can make an accurate model of
the inverse-Compton emissivity in the TeV regime.

Detailed inverse-Compton modelling of kpc-scale jets is challenging
because of the large number of photon fields that need to be taken
into account (as discussed above) and because of the strong
point-to-point variation in the photon energy density in several of
them; it is clear that the traditional methods of constructing
one-zone synchrotron/inverse-Compton models are unlikely to give
answers that can be relied on to better than an order of magnitude. In
this paper we present a framework for inverse-Compton modelling of M87
and Cen A and show that the existing TeV detections already provide
quite strong constraints on the magnetic field strength in these
objects. We also discuss the prospects for better constraints on these
objects, and perhaps detections of others, with the next-generation
capabilities to be provided by the Cerenkov Telescope Array (CTA).

The remainder of the paper is organized as follows. In Section
\ref{code} we provide an overview of the code that we use to carry out
the jet modelling and inverse-Compton calculations, and in Section
\ref{limitations} we comment on some of the limitations imposed by our
approach. Section \ref{models} describes the general approach we take
to matching our jet models to the data and the specific constraints
that we use for Cen A and M87. Section \ref{results} gives the results
of our modelling, Section \ref{outlook} describes our view of the
outlook for future facilities, and conclusions are given in Section
\ref{conclusions}.

Throughout the paper we assume a distance of 3.7 Mpc to Cen A (the
average of 5 distance indicators presented by Ferrarese \etal\ 2007; a
slightly higher distance, 3.8 Mpc, is suggested by later work, e.g.
Harris \etal\ 2010, but we retain the lower distance for consistency
with our earlier work, noting that it makes no significant difference
to our results)
and a distance of 16.4 Mpc to M87 (a weighted mean of 4 estimators
presented by Bird \etal\ 2010). For objects at larger distances we
assume $H_0 = 70$ km s$^{-1}$ Mpc$^{-1}$, $\Omega_M = 0.3$, and
$\Omega_\Lambda = 0.7$.

\section{The code}
\label{code}

The code we use is based on the framework described by Hardcastle
\etal\ (2002: hereafter H02). In that paper we were concerned
primarily with the X-ray synchrotron-self-Compton emission from the
bright, spatially resolved hotspots of FRII radio galaxies: the new
feature of the code described there was the fact that it allowed us to
consider any spatial distribution of the scattering electrons, rather
than, as previously, restricting ourselves to uniformly filled
spherical regions. Such an approach is also required for our current
problem since, as discussed above, the photon fields illuminating the
jet are strongly dependent on position, so that one-zone models will
be completely inadequate. Because the use of this code is the main new
feature of the paper, in this section we describe it in detail.

The basic principle of all parts of the code is the idea that
spatially resolved structures can be modelled in terms of
three-dimensional Cartesian `grids' of finite-sized, cuboidal volume
elements, with each volume element constituting a separate `zone' of
the synchrotron and inverse-Compton modelling and having one or more
rest-frame physical values associated with it. So, for example, the
synchrotron emission from the jet is modelled by two such grids, one
containing values for the electron energy spectrum normalization and
one for the magnetic field strength. (A third grid, used by H02 but
not in the present work for reasons that will be discussed below,
specifies which of a finite number of electron spectral models
describe each region.) Computing a quantity such as the total
synchrotron flux density from the source then involves doing a
separate synchrotron emission calculation for each non-zero element of
the grid and generating another grid, in this case one of synchrotron
volume emissivity at a specified frequency, which can then be summed
over to give the luminosity or flux density from the modelled object
(transformed into the observer's frame).
In addition, importantly, the synchrotron emissivity grid can be
projected on the plane of the sky to give an image that is
proportional to the observed synchrotron surface brightness. The main
operations of the code can all be described in terms of operating on a
parameter file and, ordinarily, one or more grids and generating a new
grid, which may then be operated on in turn.

This structure means that, in order to compute inverse-Compton
properties of the modelled source, we need to compute the
inverse-Compton emissivity for each element of the grid separately.
Doing this, of course, is what gives us the ability to make models of
the source with adequate resolution to represent its real physical
behaviour, but it comes at the cost of greatly increased computational
requirements, since to achieve adequate resolution (which in practice
means having a cell size significantly smaller than the jet radius and
much smaller than the characteristic scale of the stellar emissivity
profile) our grid needs to have $\sim 10^4$ non-zero volume elements
distributed throughout the jet. The inverse-Compton kernel that we use
is the one presented in great detail by Brunetti (2000): while in H02
we used the approximations given by Brunetti appropriate for Thomson
scattering in the rest frame, in this paper we use the full version
with no approximations, so as to take account of Klein-Nishina
effects, which adds to the computational complexity of the problem.
Specifically, our inverse-Compton emissivity, assuming an isotropic
electron population within each cell, is given by Brunetti's equations
31 and 32, and we find the minimum Lorentz factor capable of
scattering between two given energies using equations 33 and 13. The
emissivity is then integrated over all available photon and electron
energies. Since jets are beamed, we also need to take account of the
effects of special relativity in the inverse-Compton calculations. For
lab-frame isotropic photon fields such as the CMB or the EBL, we can
simply integrate over the whole sky, taking account of the Doppler
shift and transforming all angles into the jet rest frame, to obtain a
conversion between the normalization of the electron energy spectrum
in a given cell and the inverse-Compton volume emissivity. However, in
the cases of SSC and inverse-Compton scattering of starlight, where
the photon field is anisotropic, the problem is harder: we need to
compute the illumination of every grid element by a large number of
other grid elements, and in principle each one has a different Doppler
factor and a different jet-frame inverse-Compton scattering angle. We
use a modified version of the simplification described by H02, in
which we first calculate the integral of Brunetti's eq.\ 31 over a
reasonably well-sampled lookup table of Doppler factors and scattering
angles and then interpolate over this lookup table to find the
contribution to the inverse-Compton emissivity for every illuminating
cell and for every illuminated cell. While this process is still
computationally expensive, it is very much less so than it would be if
we were to do the full integral for every cell, although it does
require the use of some simplifying assumptions, which are described
in more detail below. The SSC and starlight-illumination codes make
use of the MPI framework, as implemented in MPICH2\footnote{See
  http://www.mcs.anl.gov/research/projects/mpich2/}, to allow the
calculation of the lookup table and the actual emissivity computation
to be distributed over a large number of machines: while individual
calculations are practical on modern multi-core desktop machines, the
final computations in which jet speed and angle to the line of sight
were allowed to vary were done on the University of Hertfordshire
cluster\footnote{http://star.herts.ac.uk/progs/computing.html},
typically using 16 physical compute servers and 128 cores.

\section{Limitations and approximations}
\label{limitations} 

The models we use have the following restrictions and limitations.

\begin{enumerate}
\item We assume a single electron energy spectrum throughout the jets.
  Allowing for a discrete number of different spectra, as in H02, or,
  worse, a continuously spatially varying electron spectrum throughout
  the jet would prevent us from applying the computational
  simplification described above -- in effect, we would have to do the
  full numerical integration for the illumination of every cell by
  every other cell in the SSC and starlight cases. At present this is
  computationally intractable. Since we know that the synchrotron
  spectra of the jets actually do vary as a function of position, at
  least at high energies (e.g.\ Hardcastle \etal\ 2007) this is not
  an ideal approximation, but it is necessary to make progress.
\item We assume a single value of the magnetic field strength
  throughout the jet. This is also needed to simplify the synchrotron
  self-Compton calculations; without it, the illuminating synchrotron
  spectrum from each region of the jet would be different, so again we
  would greatly increase the required number of integrals over the
  Brunetti equations. This constraint coupled with the previous one
  means (a) that the only means we have of modelling the surface
  brightness variation of the jet is to vary the {\it normalization}
  of the electron energy spectrum, and (b) means that we make the
  implicit assumption that the jet looks the same at all frequencies
  -- which again we know to be an approximation that breaks down at
  the highest energies. Because the equipartition field strength
  depends only weakly on electron energy density, this approximation
  is not likely to be seriously problematic for realistic jet models.
\item We assume a uniform velocity field throughout the jet -- all
  components of the jet are moving at the same speed and the same
  direction. We know from modelling such as that of Laing \& Bridle
  (2002) that this is not likely to be accurate, and that the
  magnitude and direction of the velocity field are likely to vary
  radially and along the jet. In fact,
  incorporating a varying velocity field into our modelling would not
  be too difficult, but we have little or no constraint on anything
  but the bulk jet speeds for our targets, and so have not implemented
  this feature. The effects of beaming on our results are in fact
  relatively modest, given the jets' low speeds and comparatively
  large angles to the line of sight (see discussion of M87 below,
  Section \ref{m87-speed}), and so we do not expect this
  assumption to have any very important effects.
\item We assume that the starlight volume emissivity distribution is
  spherically symmetrical and can be modelled by a Mellier \& Mathez
  (1987) distribution, based on the deprojection of a de Vaucouleurs
  ($r^{1/4}$) profile.
  More importantly, we assume, for the same reasons as for the
  previous point, that the emission spectrum from each region of the
  host galaxy is the same. In the case of M87, this is not
  too bad an approximation; in the case of Cen A, we know that the
  dust emissivity must in reality be very differently distributed from
  that of the starlight. However, both are strongly centrally peaked,
  which is the most important factor from the point of view of their
  effect and the jet, and we have not considered it necessary, for
  example, to make a two-component model of the dust and starlight
  emissivity in Cen A, although the structure of the code means that
  that would be possible in future if a more detailed model were warranted.
\item We neglect any illumination from the hidden AGN or blazar.
  Self-illumination of the modelled part of the jet 
  is obviously taken account of by the SSC modelling, but we have very
  little information on the structure and intrinsic luminosity of the
  pc-scale jet and even less on the spectrum of the AGN as seen by the
  kpc-scale jet. This in principle means that any flux/luminosity
  calculations we consider should probably be taken to be lower
  limits. In practice we may not be making too large an error by
  neglecting these photon fields: both are Doppler-suppressed in the
  frame of the kpc-scale jet and both suffer strongly from
  inverse-square dilution with distance from the centre of the galaxy,
  which is not the case for the starlight or SSC fields.
\item We neglect photons from the diffuse, kpc-scale lobes of our
  targets: these are not likely to be energetically important compared
  to the other photon fields and the comparatively low-frequency
  photons they predominantly emit will be hard to scatter to TeV
  energies.
\item Finally, we do not attempt to model the attenuation of TeV
  $\gamma$-rays by photon-photon interactions within the host galaxy
  (our objects of interest are close enough that interactions with the
  EBL as the $\gamma$-rays travel between the sources and the detector
  may safely be neglected). The cross-section for interactions of TeV
  photons peaks at target photon energies corresponding to the
  near-IR, where the photon density is relatively low, and modelling
  has shown that attenuation by host-galaxy starlight can be safely
  neglected both for the Milky Way (Moskalenko, Porter \& Strong 2006)
  and for more strongly star-forming galaxies (Gilmore \& Ramirez-Ruiz
  2010). In the specific case of Cen A, Stawarz \etal\ (2006a) have
  shown that the attenuation is at the $\sim 1$ per cent level. We
  therefore feel justified in neglecting this effect. 
\end{enumerate}

\section{Modelling the jets and host galaxies}
\label{models}

Given the constraints above, our modelling of the jets involves
attempting to reproduce (1) the observed {\it overall} jet synchrotron
spectral energy distributions and (2) the approximate appearance of
the jets in radio synchrotron emission. Because we know that
high-energy electrons are necessary to produce high-energy photons, we
only model in this way the region of the jets that are detected in
X-rays.

We know that the integrated spectra of these jets tend to be
adequately modelled as broken power-laws (Hardcastle \etal\ 2001,
2006) and so objective (1) is achieved by determining the low-energy
electron energy index from observations in the radio through optical
band, by taking the X-ray spectral index to indicate the high-energy
electron energy index and by adjusting the overall normalization and
the energy of the break, $\gamma_{\rm break}$, so as to ensure that
the synchrotron spectrum for the jet always passes through the radio
and X-ray data points. At the same time, we adjust the magnetic field
strength in the jet so as to obtain {\it overall} equipartition --
that is, the total energy in electrons and magnetic field integrated
over the jet are equal. This does not correspond to equipartition at
every point in the jet because, as discussed above, the electron
spectrum normalization varies from point to point in the grid, while
there is only a single magnetic field strength throughout the jet.
However, no point in the jet is very far from this more traditional
equipartition condition. We assume a low-energy cutoff $\gamma_{\rm
  min} = 100$ and a high-energy cutoff $\gamma_{\rm max} = 5 \times
10^9$ (i.e. corresponding to electrons with PeV energies, well above the
maximum energy to which Cerenkov telescopes are sensitive; the results
are not sensitive to this value as long as it is above $\sim 10$ TeV).

Objective (2) is achieved by adjusting the spatial dependence of the
electron normalization so as to give a reasonable match to the
observed synchrotron surface brightness. We model regions of the jets
as simple geometrical structures, such as cylinders or truncated
cones, with uniform or smoothly varying electron densities as a
function of position within the structure. Because we have to model
the three-dimensional electron distribution, we do not attempt to
reproduce the detailed structures seen in either the radio or the
X-ray synchrotron emission, and of course (as discussed above) we
cannot attempt to reproduce the differences between them. Our models
are thus really intended to represent the jets rather than to give a
detailed picture. We emphasise though that the structures that we do
not model (principally compact knots seen predominantly in X-ray) are
not expected to make a very strong contribution to the inverse-Compton
flux from the sources.

A key factor in the modelling of both jets is the jet bulk Lorentz
factor and angle to the line of sight. Obviously these have a strong
effect on the rest-frame energy density in both electrons and magnetic
field, but also photons: the latter effect gives rise to the
well-known boosting of inverse-Compton scattering of isotropic photon
fields like the CMB in the case of highly beamed jets (e.g. Tavecchio
\etal\ 2000) but the former must also be accounted for, in the sense
that the observed radio and X-ray data points must be corrected for
Doppler effects before the electron energy normalization and magnetic
field strengths are corrected. Slightly less obviously, the unknown
angle to the line of sight has a strong effect on the volume of the
jet (also affecting the magnetic field strength and electron energy
density) and the spatial positions it occupies (which is important in
the cases of SSC and inverse-Compton scattering of starlight, where
the photon field depends on position).

The major direct constraint on the bulk Lorentz factor $\Gamma$ and
the angle to the line of sight $\theta$ comes from observations of
proper motions in the kpc-scale jets (Biretta, Zhou \& Owen 1995; Hardcastle
\etal\ 2003; Goodger \etal\ 2010). If we assume that the observed
speeds correspond to bulk motions, then the apparent speeds
$\beta_{\rm app}$ are related to the true underlying $\beta$ by the
standard formula
\[
\beta_{\rm app} = \frac{\beta \sin \theta}{1-\beta \cos \theta}
\]
So for any choice of angle to the line of sight, we can compute a
corresponding $\beta$ from the measured value of $\beta_{\rm app}$.
Our initial approach in jet modelling was therefore to use a fixed
value of $\beta_{\rm app}$, determined from observations, and model
the jet at all angles to the line of sight, computing the
inverse-Compton flux density at 1 TeV for all photon fields. This
gives us a sense of the effect of this angle uncertainty on our TeV
predictions. We then focus on the most likely angle to the line of
sight and on more plausible beaming models in the later parts of the
paper.

The following subsections give details of the models applied to each
of the two jets.

\subsection{Cen A}
\label{cena-model}

\begin{figure*}
\epsfxsize 14cm
\epsfbox{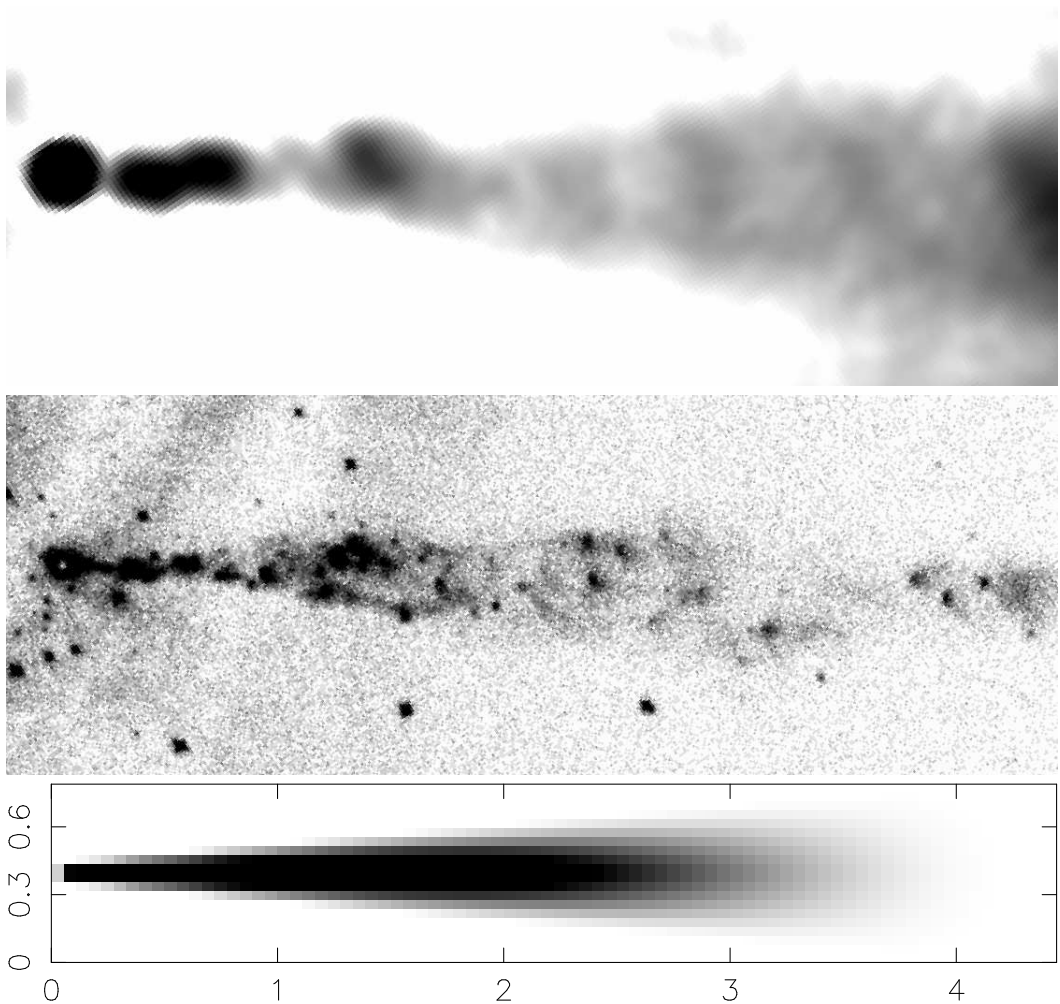}
\epsfxsize 14cm
\epsfbox{cena_smodel.ps}
\caption{Comparison of the real and model synchrotron images of Cen A.
  Top: 5-GHz VLA map with 6-arcsec resolution showing the radio
  emission from the X-ray bright region of the jet. Middle: {\it
    Chandra} X-ray image from the data used by Hardcastle
  \etal\ (2007). Bottom: model synchrotron emission, assuming
  $\theta=50^\circ$; axis labels are in (projected) kpc. The model does not
  attempt to reproduce the knotty and patchy structure seen in the
  real jet (the inner part of the jet is dominated by knots in the
  radio and the X-ray) but does try to reproduce the general run of
  surface brightness in the diffuse X-ray emission. The core (point source to
  the left, top and middle panels) is not represented on the model
  image.}
\label{cena-synchrotron}
\end{figure*}

The radio flux density of the Cen A jet, 30.6 Jy, is measured from the
1.4-GHz map of Hardcastle \etal\ (2006). The X-ray flux
density at 1 keV, 134 nJy, is the sum of the normalizations of the {\it
  extended} emission regions considered by Hardcastle \etal\ (2007).
We do not consider the compact knots, although, as shown by Goodger
\etal\ (2010), these account for about half the absorption-corrected
1-keV flux density of the source. In general compact features like
these knots are rather poor sources of inverse-Compton emission where
scattering of external photon fields is concerned, though they are
better sources of SSC emission. In any case, we lack the detailed
information on the knots' structure and electron energy spectrum
needed to model them in the framework provided by our code, so we have
chosen to model only the extended component. This means that our
predictions for SSC emission from Cen A are probably on the
conservative side.

We model the electron energy spectrum using the model fitted to the
radio, mid-IR and X-ray data by Hardcastle \etal\ (2006), which is a
broken power-law in electron energy: the low-energy electron energy
index is 2.06 steepening to 3.88 at high energies, corresponding to
the photon index of 2.44 measured by Hardcastle \etal\ (2006) for the
middle region of the X-ray jet. As described above, the normalization
and break energy of the electron spectrum are adjusted to keep
consistency with the observed radio and X-ray measurements. We now
know (Hardcastle \etal\ 2007) that the photon index even of the
extended component of the jet, as well as the radio/X-ray ratio,
varies with distance along the jet, but, as discussed above, we need
to adopt a single electron spectrum in order to make progress.

\begin{figure*}
\epsfxsize 14cm
\epsfbox{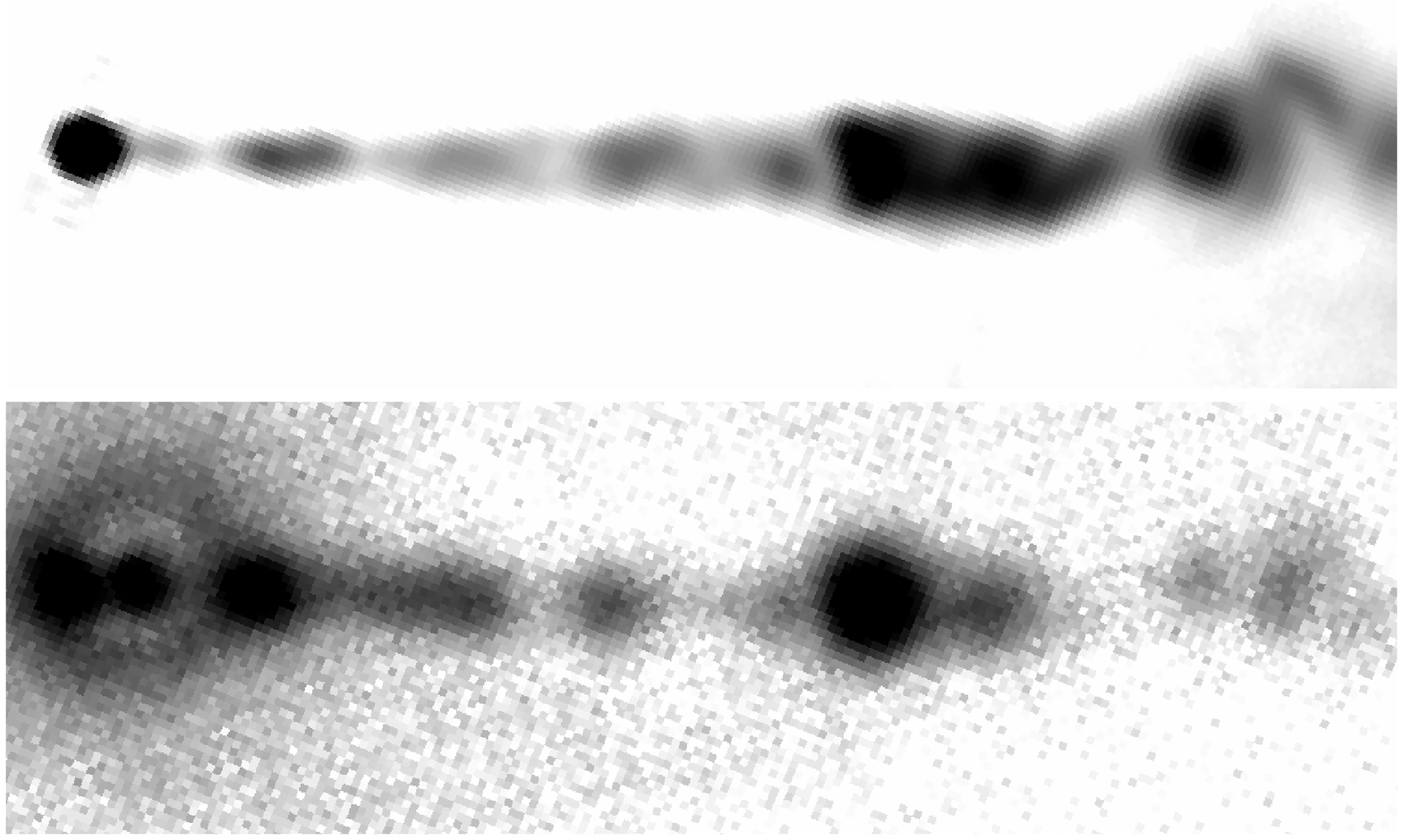}
\epsfxsize 14cm
\epsfbox{m87_smodel.ps}
\caption{Comparison of the real and model synchrotron images of M87.
  Top: 5-GHz VLA map with 0.4-arcsec resolution showing the radio
  emission from the X-ray bright region of the jet. Middle: {\it
    Chandra} X-ray emission from the same region. Bottom: model
  synchrotron emission, assuming $\theta=45^\circ$; axis labels are in
  kpc. The core is not represented in the model image.}
\label{m87-synchrotron}
\end{figure*}
We model the spatial structure of the jet as a truncated cone with
inner radius 3 arcsec, outer radius 22 arcsec, and projected length
230 arcsec. Within this cone, the electron normalization decreases
linearly with radius and as $(1-d)^2$ where $d$ is the distance along
the axis of the cone. This electron distribution gives a reasonable
match to the observed (knot-free) X-ray surface brightness
(Fig.\ \ref{cena-synchrotron}): we constrain our models to reproduce
the spatial distribution of the diffuse X-ray
synchrotron emission given that we are most interested in the
distribution of the TeV electrons. The results of our modelling are
only weakly sensitive to our assumptions about the spatial
distribution of the electrons.

We take $\beta_{\rm app}$ to be 0.534, as measured by Goodger
\etal\ (2010) for knot A1B, and for the host galaxy modelling we adopt
the values used by Croston \etal\ (2009); flux densities for NGC\,5128
from the ultra-violet down to the far-IR are taken from NED. Both for
Cen A and for M87 we use the model for the EBL used by Hardcastle
\etal\ (2009), i.e. the model of Raue \& Mazin (2008).

Finally, we take the measured VHE flux to be $(1.56 \pm 0.67) \times
10^{-12}$ photons cm$^{-2}$ s$^{-1}$ above 250 GeV, and the measured
photon index to be $2.73 \pm 0.65$, as reported by Aharonian
\etal\ (2009). This means that the 1-TeV flux density of Cen A is
$(1.6 \pm 0.7) \times 10^{-16}$ Jy, if we assume a fixed photon index
of 2.73.

\subsection{M87}
\label{m87-model}

The normalizing radio flux density of M87, 5.25 Jy, is taken from a
5-GHz VLA radio map at 0.4-arcsec resolution. The 1-keV X-ray flux
density we use (279 nJy) and the photon index (2.21) are derived from
fitting a single power-law model to 667 ks of archival {\it Chandra}
data (Harwood \& Hardcastle, in prep.). The region used for the
spectral extraction includes the whole extended X-ray jet, but
excludes the bright, strongly varying knot HST-1 (e.g. Harris
\etal\ (2006). Clearly measuring the X-ray flux in this way is less
conservative than our approach in the case of Cen A: in M87 we cannot
separate discrete X-ray knots from the diffuse X-ray emission, since
the spatial resolution is so much worse. If the situation in Cen A
were replicated in M87, then the total X-ray flux from the jet would
exceed the X-ray flux density from the diffuse component only by a
factor $\sim 2$. As with Cen A, we model the overall electron spectrum
as a broken power law, with the same injection index as for Cen A and
with a break that replicates the observed photon index in the X-ray.

We model the jet as a combination of a truncated cone of length 11
arcsec, with inner radius 0.4 arcsec and outer radius 1.0 arcsec, and
a cylinder of radius 1.0 arcsec and length 6.5 arcsec, with electron
density decreasing linearly with radius in both structures, and with a
jump in electron density of a factor 1.5 at the boundary between the
cone and the cylinder. This roughly reproduces the observed radio
structure of the M87 jet (Fig.\ \ref{m87-synchrotron}) with the
boundary between conical and cylindrical structures representing the
change in the jet opening angle and surface brightness at knot A.

Jet speeds are estimated using the observations of Biretta
\etal\ (1995). In this very detailed study various components of M87
are observed to have different apparent speeds: Biretta \etal\ in fact
suggest that the observations are consistent with high bulk speeds
($\gamma \sim 3$) in the inner jet and much slower speeds beyond knot
A, which is consistent with the interpretation of knot A as a jet-wide
shock. Modelling the SSC emission in this scenario would require us to
calculate the mutual illumination of two jet regions moving
relativistically with respect to each other, which is not yet
implemented in our inverse-Compton code. For the sake of simplicity
and for ease of comparison with Cen A we adopt $\beta_{\rm app} =
0.479$, which is the speed quoted by Biretta \etal\ for the edge of
knot A. However, we consider below the possibility that the jet may
have a bulk speed much higher than this.

We model the host galaxy in the same way as that of Cen A, assuming
that it follows a de Vaucouleurs profile, with parameters taken from
Liu \etal\ (2005). For optical flux densities to determine the
starlight SED we again take values from NED, but at IR wavelengths we
need to take into account the effects of contamination from
synchrotron emission: Baes \etal\ (2010) have used {\it Herschel} data
to put upper limits on the amount of cold dust that can be present in
the host galaxy, showing that almost all the FIR emission lies on a
simple extrapolation of the known lower-frequency synchrotron
spectrum. We base our model in the mid to far-IR on the modified
black-body model that they fit to the residual fluxes after
subtraction of the synchrotron component, which means that M87 is very
weak in this region of the spectrum compared to Cen A.

At TeV energies M87 is known to be strongly variable (Acciari
\etal\ 2009): in the past it has been argued that the flaring TeV
emission might originate in HST-1 (e.g. Cheung \etal\ 2007) but
observations of a strong flare at TeV energies that had no X-ray
counterpart but was close to the start of a jet ejection event in the
radio led Acciari \etal\ (2009) to argue that the most likely site of
the strongly varying TeV component is in the parsec-scale jet. In any
case, it is clear that any component of the TeV emission variable on
timescales of days to weeks cannot be related to the diffuse emission
from the kpc-scale jet. For comparison with our model predictions we
have taken the average flux density measured by Acciari \etal\ (2010)
from the VERITAS observations in 2009, in which no significant
variability of the TeV emission was observed. They report a flux of
$(1.59 \pm 0.39) \times 10^{-12}$ photons cm$^{-2}$ s$^{-1}$ above 250
GeV with $\Gamma = 2.5$ (as measured for the 2008 data, which however
include a gamma-ray flare), which is coincidentally almost identical
to what we have used for Cen A above, and which corresponds to a 1-TeV
flux density (assuming $\Gamma = 2.5$) of $(2.0 \pm 0.49) \times
10^{-16}$ Jy.

\section{Results}
\label{results}

\subsection{Cen A}

\begin{figure*}
\epsfxsize 16cm
\epsfbox{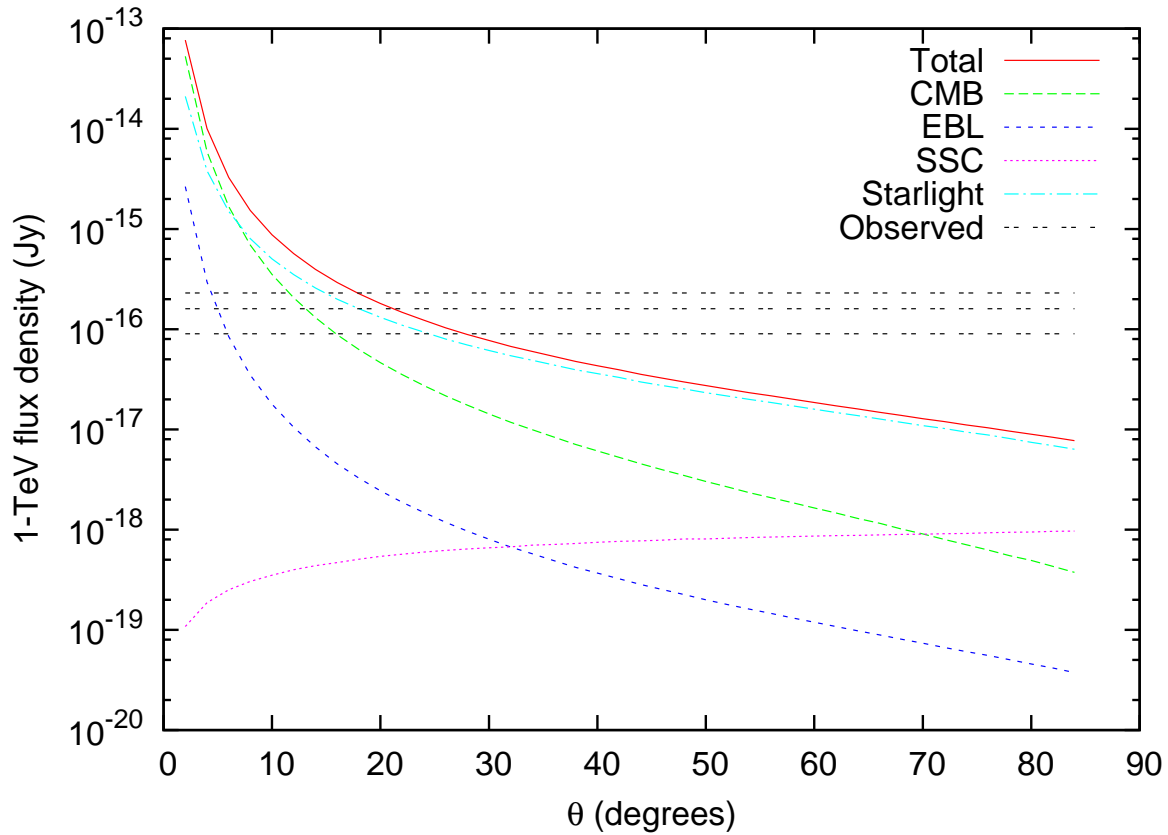}
\caption{Predicted 1-TeV flux density from the Cen A jet as a function
  of the angle made by the jet to the line of sight, $\theta$. The
  three dashed horizontal lines show the best estimates of the 1-TeV flux
  density from the HESS measurements and its $1\sigma$ error.}
\label{cena-tev-theta}
\end{figure*}

Fig.\ \ref{cena-tev-theta} shows the predicted 1-TeV inverse-Compton flux
density as a function of jet angle to the line of sight $\theta$ for
Cen A, given the model discussed in Section \ref{cena-model}, for an
equipartition $B$-field as discussed in Section \ref{models}.

The general features of this plot are as expected from first
principles. SSC emission is strongest when the jet is in the plane of
the sky, mainly because it is physically smallest in that orientation
so that the inverse-square suppression of SSC is least, but also
because the correction for Doppler suppression increases both the
electron and the photon density in the rest frame of the jet. CMB and
EBL flux densities both increase as the angle to the line of sight
gets smaller, both because of the increasing volume of the jet, which
means that the total number of scattering electrons increases, and, at
small angles, because of the effects of Doppler-boosting of these
isotropic photon fields into the jet rest frame. The starlight
component scales in a similar way, but increases more weakly at very
small angles to the line of sight because for these angles the bulk of
the jet is well away from the centre of the galaxy where the energy
density in starlight is at its peak, and because for these angles most
of the scattering, especially when beaming effects are taken into
account, is at small deflection angles where the cross-section for
inverse-Compton scattering is lowest. We see that the predicted 1-TeV
flux density is very similar to what is measured using HESS for $20 <
\theta < 30^\circ$; at smaller angles to the line of sight we
overpredict the TeV flux density.

\begin{figure}
\epsfxsize 8.5cm
\epsfbox{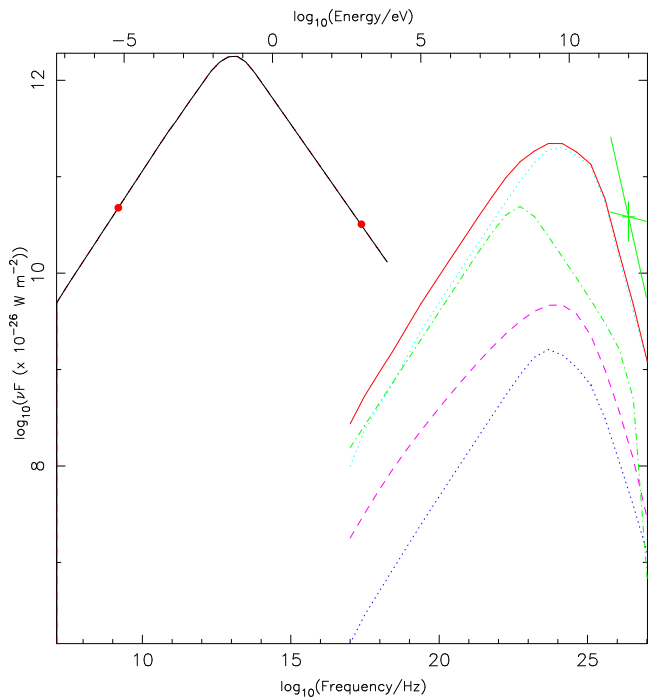}
\caption{Spectral energy distribution for the inverse-Compton emission
  for a model of the Cen A jet with $\theta = 50^\circ$, $\beta = 0.51$ and
$B=B_{\rm eq}$. The solid black line is the synchrotron emission, and
  the solid red line the total inverse-Compton from all modelled
  photon fields. The separate contributions from the four photon
  fields are also plotted: from lowest to highest at 1 TeV, these are
  the EBL (orange line), the CMB (blue line), SSC (green line) and
  starlight (magenta line). The red dots show our constraints on the
  radio and X-ray flux densities and the green cross is our adopted
  1-TeV flux density, with the bow-tie showing the observational
  uncertainties on the photon index. Frequencies and energies are
  plotted in the observer's frame.}
\label{cena-spec}
\end{figure}

The actual angle to the line of sight for Cen A is not well known:
Hardcastle \etal\ (2003) have summarized the constraints from radio
data. If we adopt their favoured value of $50^\circ$, we see that the
inverse-Compton prediction is $0.27 \times 10^{-16}$ Jy, a factor
$\sim 6$ below the flux density measured by HESS.
Fig.\ \ref{cena-spec} shows the predicted inverse-Compton SED from
such a model. We see that the spectral index in VHE $\gamma$-rays is
steeper than that in the X-ray, which is a result of the Klein-Nishina
correction to the scattering cross-section: the predicted photon index
for the total inverse-Compton spectrum is in fact around 3.1, but this
is consistent within the errors with what is observed by HESS (note
that if we assume that this is the correct spectral index, the 1-TeV
flux density calculated from the HESS measurement comes down to $(1.2
\pm 0.5) \times 10^{-16}$ Jy, a factor $\sim 4$ above the equipartition
prediction). We conclude that for equipartition and for our best
estimates of the jet properties ($\theta =
50^\circ$, $\beta = 0.51$) inverse-Compton scattering from the
kpc-scale jet is unable to produce the observed TeV emission, although
it is possible to produce all the observed $\gamma$-rays with smaller
angles to the line of sight (we note that Hardcastle \etal\ 2003 suggested that
$\theta \sim 20^\circ$ was required on the assumption of
jet/counterjet symmetry in the kpc-scale jet).

\begin{figure}
\epsfxsize 8.5cm
\epsfbox{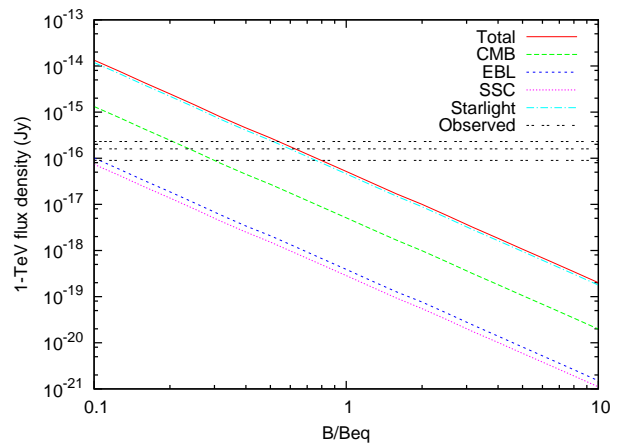}
\caption{Inverse-Compton flux density from the jet of Cen A as a
  function of magnetic field strength, normalized to the equipartition
  value, for the model with $\theta = 50^\circ$ discussed in the text.
  Magnetic fields only a small factor below the equipartition field
  can allow the model to reproduce all the observed TeV emission.}
\label{bbeq}
\end{figure}

Any departure from equipartition alters the situation significantly.
As $B$ decreases below $B_{\rm eq}$, the electron number density and
thus inverse-Compton emissivity increase rapidly (Fig.\ \ref{bbeq}).
The predicted flux goes roughly as $(B/B_{eq})^{-2.4}$ at 1 TeV: so,
for the model with $\theta = 50^\circ$ discussed above, the HESS data
require $B \ga 0.6 B_{\rm eq}$. In the lobes and hotspots of FRII
radio galaxies (e.g.\ Hardcastle \etal\ 2004, Kataoka \& Stawarz 2004,
Croston \etal\ 2005) we routinely see departures from equipartition at
this level, so we do not regard it as at all implausible that the TeV
emission really is wholly produced by the kpc-scale jet for $\theta
\sim 50^\circ$. We can also say that models where $B$ is more than a
factor of a few below $B_{\rm eq}$ are convincingly ruled out by the
data, irrespective of $\theta$, particularly when we consider that
many possible other emission processes may contribute to the observed
TeV emission (Aharonian \etal\ 2009) and that our models of the source
TeV electron population in Cen A are quite conservative. Stawarz,
Kneiske \& Kataoka (2006b) reached a similar conclusion based on the
FRI source population's contribution to the $\gamma$-ray background,
but this is the first time that TeV emission has been used directly to
constrain the magnetic field strength in an individual source.

\subsection{M87}
\label{m87-speed}

\begin{figure*}
\epsfxsize 16cm
\epsfbox{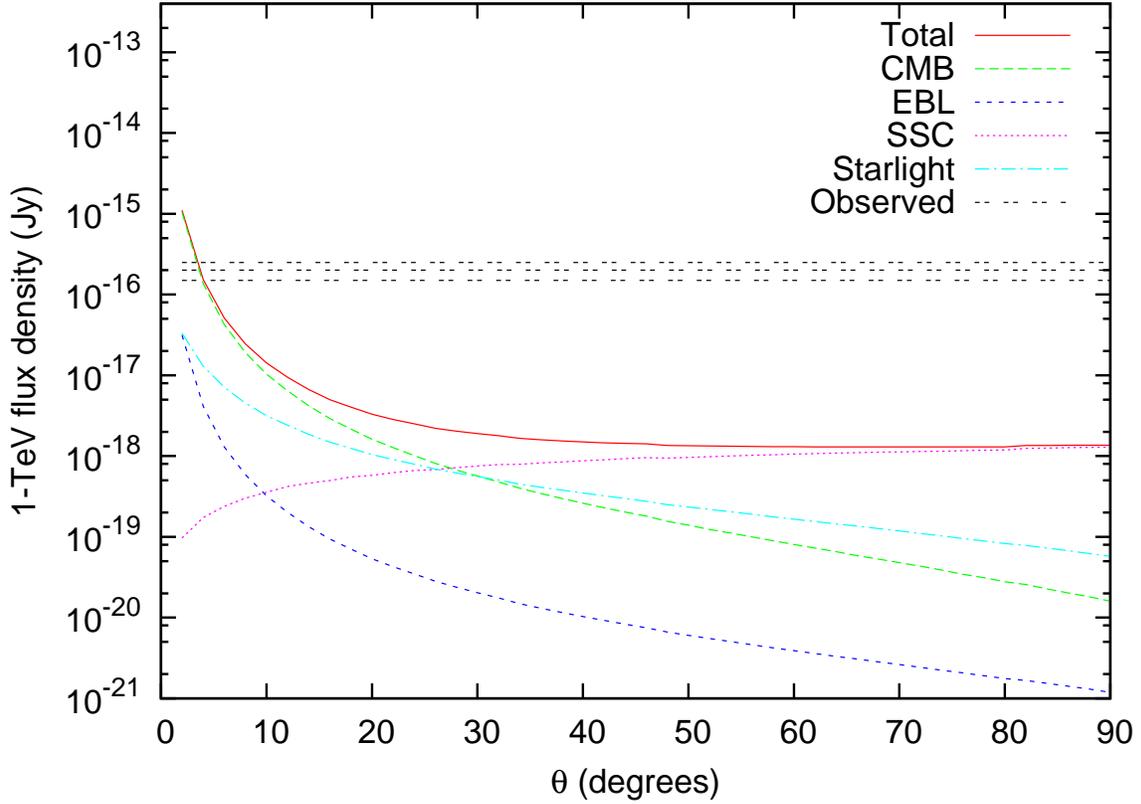}
\caption{Predicted 1-TeV flux density from the M87 jet as a function
  of the angle made by the jet to the line of sight, $\theta$. The
  three dashed horizontal lines show the best estimates of the 1-TeV
  flux density from the 2009 VERITAS measurements, as discussed in the
  text, and its $1\sigma$ error.}
\label{m87-tev-theta}
\end{figure*}

\begin{figure*}
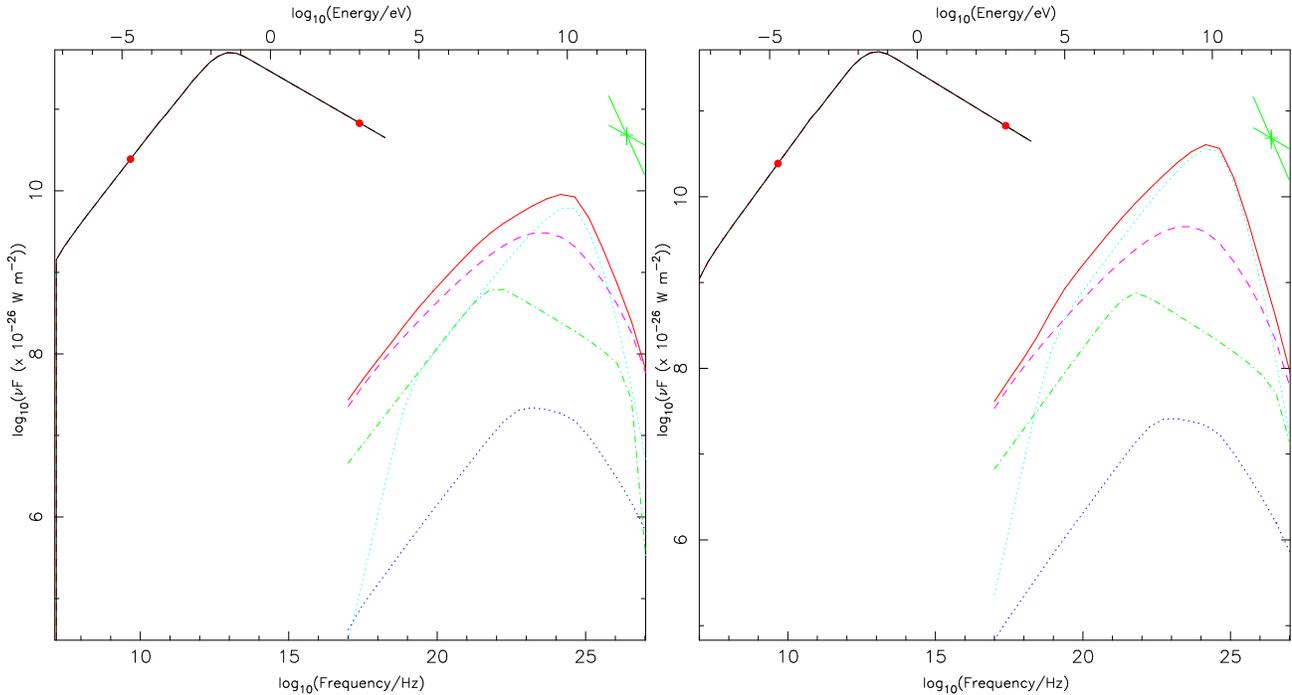

\epsfxsize 8.5cm
\epsfbox{m87_jet_ic.ps}
\epsfxsize 8.5cm
\epsfbox{m87_jet_ic-g3.ps}
\caption{Spectral energy distribution for the inverse-Compton emission
  for a model of the M87 jet with $B=B_{\rm eq}$. Left panel: $\theta
  = 45^\circ$, $\beta = 0.46$. Right panel: $\theta = 45^\circ$,
  $\gamma = 3$. Lines and points as in Fig.\ \ref{cena-spec}: the
  bowtie shows $\Gamma = 2.5 \pm 0.3$, as measured by Acciari
  \etal\ (2010) for the pre-flare data in 2008, as no measurement is
  available for the 2009 dataset.}
\label{m87-spec}
\end{figure*}

The situation is clearly different for the M87 jet.
Fig.\ \ref{m87-tev-theta} again shows inverse-Compton emission as a
function of angle to the line of sight for our adopted model with
$\beta_{\rm app} = 0.479$. In this model, SSC emission actually
dominates over the other photon fields for large $\theta$, presumably
a consequence of the fact that M87's jet is physically smaller than
and significantly more luminous than Cen A's. The key difference,
though, is that the net IC emission does not approach the levels of
the 2009 VERITAS flux for $\theta \ga 10^\circ$: for $\theta =
45^\circ$ (the SED for which is plotted in Fig.\ \ref{m87-spec}), it
is more than two orders of magnitude below the observations, and a
departure from equipartition of a factor $\sim 8$ would be required to
bring the flux up to observable levels. Even assuming $\gamma = 3$ for
the whole jet [as argued by Biretta \etal\ (1995) for the inner jet
  only], while retaining $\theta=45^\circ$, only increases the
predicted 1-TeV flux density by a factor $\sim 2$. We can again claim
to have constrained the field strength from these observations, but it
is a much weaker constraint, and our prediction is that the bulk of
the TeV emission from M87 must come from some other source unless the
angle to the line of sight is very small. One interesting point here
is the relative weakness of the starlight component in M87 compared to
that Cen A: although the energy density in starlight in M87 is
significantly higher than that in the CMB, and this is reflected in
the relative normalizations of the peaks in Fig.\ \ref{m87-spec}), the
starlight contribution actually falls to levels comparable to that of
the CMB emission by 1 TeV, whereas `starlight' dominates by a large
factor at the same energies in Cen A, as can be seen in
Fig.\ \ref{cena-spec}. We believe that the reason for this is the very
different spectra adopted for the two host galaxies. The `starlight'
model in Cen A includes a large contribution from dust emission from
the central dust lane, whereas there is no evidence for dust emission
from M87 at any significant level (Section \ref{m87-model}).
Klein-Nishina effects significantly reduce the efficiency of
scattering of the predominantly optical and near-IR photons from the
M87 stellar population into the TeV band. At these energies, a simple
comparison of the energy densities of the photon populations is not a
reliable guide to their contribution to inverse-Compton emission: this
also suggests that strong dust emission may be a pre-requisite for a
bright TeV-emitting jet.

\section{Prospects for next-generation facilities}
\label{outlook}

\begin{figure*}
\epsfxsize 16cm
\epsfbox{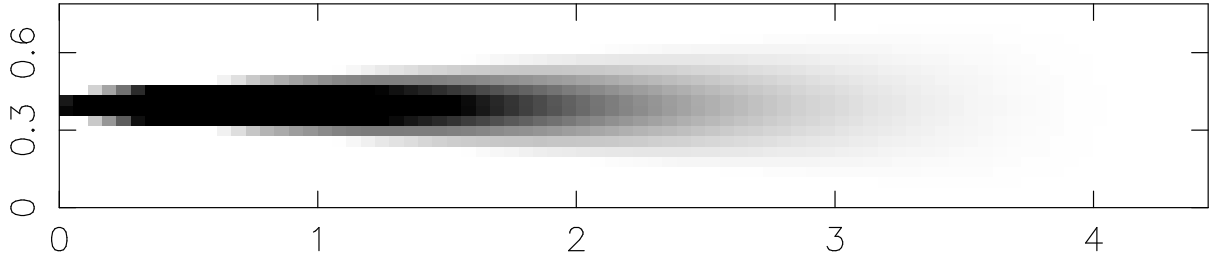}
\caption{A map of 1-TeV inverse-Compton emission from the jet of Cen A
  (assuming $\theta = 50^\circ$, $\beta = 0.51$ as described in the
  text). The image can be seen to be significantly brighter close to
  the active nucleus and the centre of the host galaxy (left-hand side) compared
  to the model synchrotron image of Fig.\ \ref{cena-synchrotron}. Axis
labels are in kpc.}
\label{cena-cta-map}
\end{figure*}

The Cerenkov Telescope Array (CTA) is the next-generation TeV
gamma-ray facility, currently in the design phase (CTA Consortium
2010). It will both be significantly more sensitive and have
significantly higher angular resolution than existing facilities such
as HESS and VERITAS. Unfortunately, the likely angular resolution
(expected to be $\la 1$ arcmin at 1 TeV: e.g. CTA Consortium 2010) is
too low to resolve even M87's kpc-scale jet (with a length in the
X-ray of $\sim 20$ arcsec) from its active nucleus, and the same is
true for most other kpc-scale X-ray jets at distances greater than
M87's. The best hope for an observational test of the idea that
kpc-scale jets can be responsible for TeV $\gamma$-ray emission will,
unsurprisingly, be provided by Cen A. Here the X-ray jet, which
extends for $\sim 4$ arcmin on the sky, should be clearly resolved by
the CTA. Unfortunately in our models, in which scattering of starlight
dominates the inverse-Compton emission, the jet is still brightest in
the unresolved central regions (Fig.\ \ref{cena-cta-map}). However, if
we take the model with $\theta = 50^\circ$ displayed in
(Fig.\ \ref{cena-cta-map}), and assume that all the currently observed
VHE emission is produced by the jet, then half the total observed
1-TeV flux density should be produced in the region outside 1 arcmin
(1 kpc), and this extended emission should be relatively easy for the
CTA to detect, given its greatly improved sensitivity over its
predecessors. A moderate-duration observation of Cen A with the CTA
will therefore very probably be able to make a conclusive measurement
of the magnetic field in an FRI jet for the first time. Even a
non-detection would place a lower limit on the magnetic field strength
that will be considerably higher than the equipartition value, and
would therefore be of great interest as no other component of a radio
galaxy for which the magnetic field strength has been measured
exhibits this kind of behaviour\footnote{There is some evidence that
  on large scales the energy density in FRI jets is dominated by
  non-radiating particles --- e.g. Croston \etal\ (2008) --- and so it
  is not completely implausible that the true magnetic field is
  greater than the field derived by assuming equipartition with the
  radiating electrons only, but without these CTA observations there
  will be no way of testing this model by direct observation.}.

Other than Cen A, the best FRI jets for study with the CTA will be the
very few more distant objects with X-ray jets that are extended on
several-arcmin -- and thus typically 100-kpc -- scales. The best
example that we are aware of is NGC\,6251 ($z=0.0243$), whose
large-scale X-ray jet, starting around 3 arcmin from the nucleus, was
first detected with {\it ROSAT} (Mack, Kerp \& Klein 1997) and later
studied in detail with {\it Chandra} and {\it XMM} (Evans
\etal\ 2005). As discussed in the latter paper, it is not even clear
whether the X-ray emission from the 100-kpc scale jet in NGC\,6251 is
synchrotron in origin, and we do not have a detailed model for the
electron spectrum or direct constraints on the jet speed or angle to
the line of sight. However, we can make some rough estimates of the
detectability of VHE $\gamma$-rays. Using the one-zone synchrotron
model presented by Evans et al, we modelled region 1 of the extended
jet as a cylinder with a linear radial dependence of electron density
and with $\theta = 45^\circ$ and $\beta = 0.7$, placed it at the
appropriate distance from the host galaxy (which we assume for
simplicity to have the same SED as M87) and computed the
inverse-Compton emission in the VHE $\gamma$-ray band. We find a 1-TeV
flux density at equipartition of $\sim 7 \times 10^{-18}$ Jy, a factor
$\sim 20$ below that measured by HESS for Cen A, and compute a
predicted photon index of 2.5. The CMB is by a large factor the
dominant photon field here -- light from the host galaxy, SSC and the
EBL all come in an order of magnitude or more lower -- and what
prevents the TeV emission being much fainter as a consequence is the
rather small break inferred by Evans \etal\ (2005) in the electron
energy spectrum for the jet, coupled with its large volume, which
means that the bulk of the jet's luminosity on these scales actually
emerges in inverse-Compton in the $\gamma$-ray regime. Although the
flux density we estimate here is only one value from the large range
that we would obtain if we had the data to carry out more detailed
modelling, the fact that it is only about an order of magnitude below
what is already measurable for Cen A is encouraging, and suggest that
NGC\,6251 and any similar objects might provide interesting targets
for long exposures with the CTA.

\section{Conclusions}
\label{conclusions}

The main conclusions from this paper can be summarized as follows:
\begin{enumerate}
\item We have developed a framework for carrying out detailed
  inverse-Compton modelling of the kpc-scale jets of radio galaxies in
  the regime where the anisotropic nature of inverse-Compton emission
  and Klein-Nishina effects are important, and applied it to the
  well-studied jets of Cen A and M87.
\item We have shown that the predicted inverse-Compton flux density in
  VHE $\gamma$-rays on plausible models of Cen A is quite comparable
  to what is observed. This means that the {\it existing} TeV
  observations of Cen A put quite strong constraints on the magnetic
  field strength in the kpc-scale jet, particularly given that we
  have not modelled either the TeV electrons present in the compact
  knots in Cen A or the photon field from the presumed hidden blazar
  in its nucleus: $B$ must be comparable to or
  larger than the equipartition field, unless our beliefs about the
  probable jet speed or angle to the line of sight are very wrong.
  This is the first time that a TeV detection of an individual source
  has been used to put constraints on the jet magnetic field strength;
  our work supports the analysis of Stawarz \etal\ (2006b), which
  was based on the $\gamma$-ray background, in suggesting that the
  magnetic field in these jets cannot be much less than the
  equipartition value.
\item Similar but weaker
  constraints can be derived for M87: it seems likely from our
  analysis that the large-scale jet contributes only a small fraction
  of the observed TeV emission from this source, which is consistent
  with the fact that the TeV emission is seen to be strongly variable.
  More observational work is needed in this case to pin down a
  non-variable component which could be compared with our predictions.
\item The prospects for study of extended X-ray jets with future, more
  sensitive VHE telescopes such as the CTA are limited primarily by
  the spatial resolution that will be achievable -- in many cases
  kpc-scale jets may be detectable but unresolved from emission
  related to the AGN or to the pc-scale jet. In Cen A, there is a very
  realistic possibility of resolving the jet, and we have shown that
  the possible 100-kpc-scale X-ray synchrotron jet in NGC\,6251, which
  should also be resolved by the CTA, may not be completely out of
  reach for such instruments. Angular resolution, rather than raw
  sensitivity, may be the limiting factor for the CTA in the field of
  extragalactic jets.
\end{enumerate}

\section*{Acknowledgements}

MJH thanks the Royal Society for a research fellowship. JHC
acknowledges support from the South-East Physics Network (SEPNet).

\end{document}